\documentclass[12pt]{article}
\pdfoutput=1
\usepackage{amsmath}
\usepackage{amssymb}
\usepackage{graphicx,bbm,mathrsfs}
\usepackage{color}

\newcommand{\lsim}{\raisebox{-0.13cm}{~\shortstack{$<$ \\[-0.07cm] $\sim$}}~} 
\newcommand{\gsim}{\raisebox{-0.13cm}{~\shortstack{$>$ \\[-0.07cm] $\sim$}}~} 
\newcommand{\beq}{\begin{eqnarray}} 
\newcommand{\eeq}{\end{eqnarray}} 

\begin{document}

\begin{flushright}LPT-Orsay--15--94, LAL-Orsay--15--427 \end{flushright} 

\vspace*{-4mm} 

\begin{center}

{\large\bf 
Diboson resonances within a custodially protected 

\vspace*{1mm} 

warped extra-dimensional scenario }

\vspace*{3mm}

{\sc Andrei~Angelescu$^1$, Abdelhak~Djouadi$^1$, Gr\'egory Moreau$^1$} 

and {\sc Fran\c{c}ois~Richard$^2$} 

\vspace{3mm}

{\small 
$^1$~ Laboratoire de Physique Th\'eorique,  CNRS and Universit\'e Paris-Sud \\  
B\^at. 210, F--91405 Orsay Cedex, France \\
$^2$~ Laboratoire de l'Acc\'el\'erateur Lin\'eaire, IN2P3/CNRS and Universit\'e Paris-Sud\\  B.P. 34,  F--91898 Orsay Cedex, France \\
}

\end{center}

\vspace*{-2mm}

\begin{abstract}
We propose an interpretation of the diboson excess recently observed by the ATLAS and  CMS collaborations in terms of Kaluza--Klein excitations of electroweak gauge bosons stemming from a realization of a warped extra--dimensional model that is protected by a custodial symmetry. Besides accounting for the LHC diboson data, this scenario also leads to an explanation of the anomalies that have been observed in the measurements of the forward--backward asymmetries for bottom quarks at LEP and top quarks at the Tevatron.  
\end{abstract}

\subsection*{1. Introduction}  \label{se:intro}

Compared to the expectations in the context of the Standard Model (SM), one of the very few anomalies that have been observed at the earlier run of the LHC with an energy up to 8 TeV and a total luminosity of about 25 fb$^{-1}$ are the excesses in the diboson spectra observed by the ATLAS \cite{ATLAS-excess} and, to a lesser degree, by the  CMS \cite{CMS-excess,CMS-WH} collaborations. In the former case, searches have been performed for di--electroweak  gauge bosons, $pp \to VV$ with $V=W/Z$, that are hadronically decaying and identified through jet--substructure techniques \cite{jet-substructure}. At a diboson invariant mass of about 2 TeV, an excess compared to the SM prediction has been observed by ATLAS \cite{ATLAS-excess} in all the detection modes $WW,WZ$ and $ZZ$ with a statistical significance of $\approx 2.5$--3$\sigma$ in each channel. Excesses in the same channels and at the same invariant mass have also been observed by the CMS collaboration \cite{CMS-excess} but with a smaller significance. In addition, CMS searched for heavy vector resonances decaying into $W$ and Higgs bosons that lead to $\ell \nu b\bar b$ final states and observed a $2.5 \sigma$ excess also at an invariant mass of approximately 2 TeV \cite{CMS-WH}. 
 
Besides the likely possibility that they are simply statistical fluctuations which will disappear with more data, these excesses in the structure of the diboson mass spectra can have several interpretations in terms of new physics and a vast literature has already appeared on the subject \cite{first,all-papers}. The most advocated and robust scenario is the production of new spin--one resonances that subsequently decay into two SM bosons. 

In the present paper, we consider an interpretation of this excess in the context of the warped extra--dimensional model proposed  by  Randall and Sundrum \cite{RS} and in which a bulk gauge custodial symmetry is introduced in order to protect the electroweak observables from large radiative corrections \cite{Custo}. In such models, the symmetry group is ${\rm SU(2)_R \times SU(2)_L \times U(1)_X}$ and there are extra weak gauge bosons $W',Z'$ in addition to the Kaluza--Klein (KK) excitations of all   states, not only of the weak bosons, but also  the photon, the gluon and the graviton, as well as for the fermions. If the heavier SM  $Q=t,b$ quarks can be localized towards the so-called TeV-brane where the Higgs boson is confined, large wave function overlaps between these fermions and the Kaluza-Klein excitations of gauge bosons can be generated and would lead to significant changes of the $VQ\bar Q$ couplings \cite{KK-cplgs}. These non--standard couplings could explain the puzzles in the forward--backward asymmetries for bottom quark production $A_{FB}^b$ as measured at LEP and for top quark pair production $A_{FB}^t$ as more recently observed at the Tevatron \cite{PDG}.  
 
The new gauge bosons can have masses in the few TeV range and can decay not only into the generally dominant $t\bar t, b \bar b$ and/or $bt$ modes \cite{KK-tt} but also into $VV$ and $VH$ diboson final states \cite{KK-VV}. In fact, such a configuration has already been predicted at the LHC in Ref.~\cite{first} where the diboson signal of a $1.5$ TeV resonance in a warped extra-dimensional context that could explain the $A_{FB}^b$ discrepancy at LEP \cite{AFBb} and the $A_{FB}^t$ anomaly at the Tevatron \cite{AFBt} has been put forward. In the present note, we will update the latter analysis and adjust it in order to comply as much as possible with the ATLAS and CMS data collected at the previous run \cite{ATLAS-excess,CMS-excess,CMS-WH}. Besides the $A_{FB}^Q$ asymmetries,  we will also discuss the compatibility of such an interpretation with the constraints set by the electroweak precision data \cite{PDG} and by the LHC Higgs measurements \cite{HiggsCombo}.

\subsection*{2. Synopsis of the model}  \label{se:setup}

The warped extra–dimensional scenario proposed by Randall and Sundrum is a 
particularly attractive extension of the SM as it provides a solution to the 
gauge hierarchy problem. Indeed, within the  model, the effective gravity scale 
is not the usual Planck mass $M_P= 2.44 \times 10^{18}$ GeV but the scale on the 
TeV brane which is suppressed by a warp factor, $M_\star= e^{-\pi k R_c} M_P$, 
where $1/k$ is the  curvature radius  of  the AdS space  and $R_c$ the  
compactification  radius. For  a  product $\pi k R_C \approx 33$, one obtains 
a fundamental scale that is close to the electroweak scale, $M_\star = {\cal 
O}(1)$ TeV. The first Kaluza–Klein excitations of the SM gauge bosons have  
approximately a common mass given by $M_{KK}= 2.45 k e^{-\pi k R_c} \approx M_\star$.   

Unfortunately, the high precision electroweak data lead a severe bound on the two scales, $M_\star \approx M_{KK} \gsim 10$ TeV, that is  not acceptable if the hierarchy problem is to be addressed. It has been shown that if the SM gauge symmetry is enhanced to the left–right custodial structure ${\rm SU(2)_L \times SU(2)_R \times U(1)_X}$ in the bulk, the data can be fitted while keeping the mass $M_{KK}$ at a few TeV. The SM gauge group  is  recovered  after  the  breaking  of  ${\rm SU(2)_R}$ into ${\rm U(1)_R}$ and ${\rm U(1)_R \times U(1)_X}$
into ${\rm U(1)_Y}$. The linear combination of the neutral fields that is  orthogonal to the hypercharge field will be the $Z'$ boson. Both the $Z'$ and the $W'$ bosons, the charged states of the ${\rm SU(2)_R}$ group, have no zero modes and their first KK excitations have masses that are very close to $M_{KK}$, $M_{V'}= 2.40 k e^{-\pi k R_c}$. 

Besides providing a solution to the gauge hierarchy problem, the version of the RS scenario with bulk matter allows for a new interpretation of the fermion mass hierarchies based on specific localizations of the fermion wave functions along the warped extra dimension \cite{RSloc}. Indeed, if the fermions are placed differently along the extra dimension, the observed patterns among the effective four-dimensional Yukawa couplings are generated as a result of their various wave function overlapping with the Higgs field, which remains confined on the so-called TeV brane for its mass to be protected. A parameter denoted $c_f k$ quantifies the five-dimensional mass attributed to each fermion and fixes its localization with respect to the TeV brane. With decreasing $c_f$, the zero mode fermions become  increasingly closer to the TeV--brane and acquire larger masses. 

Hence,  the third generation fermions interact more strongly with the  gauge bosons KK excitations as a result of the large overlap between their wave functions near the TeV brane. The heavy $t,b$ quarks are thus expected to be most sensitive to new physics effects. For instance, the couplings of the $b$--quarks to the $Z$ boson, which mixes with the neutral KK gauge boson excitations, can be altered as to solve \cite{AFBb} the longstanding anomaly observed at the LEP collider in the $Z \to b\bar b$ forward–backward asymmetry, the only high-energy observable that significantly deviates from the SM prediction \cite{PDG}. The same occurs for the top quark and, for  $M_{KK} \approx 2$ TeV, the KK gluons would contribute 
to top quark pair production at the Tevatron and could explain \cite{AFBt} the anomaly observed in its forward--backward asymmetry at high invariant masses.   

In the present study, we will consider only the first KK excitations of the various states and we will denote them simply by $A_{KK}, Z_{KK}, W_{KK}$ and $g_{KK}$ for the photon, the weak gauge bosons and the gluon. The first KK excitations of the gauge bosons from the additional gauge group, which have no zero modes, will be denoted by $W'$ and $Z'$. We assume a relatively low KK mass scale which approximately corresponds to the masses of the first $KK$ excitations, $M_{KK}= M_{A_{KK}} \approx M_{Z_{KK}} \approx M_{W_{KK}} \approx M_{g_{KK}}$. In the fermionic sector, we denote the first KK resonances of the $t,b$ quarks and the additional partners of the heavy quarks (the so-called custodians) collectively by $t'$ and $b'$. We will chose the $c_f$ parameters, and more precisely the charges $Q_V(c_f)$ or  the effective $V_{KK} f \bar f$ couplings relative to the $Vf\bar f$ ones,  in such a way that the experimental data for the forward--backward asymmetries $A_{FB}^b$ at LEP and $A_{FB}^t$ at the Tevatron are approximately reproduced, without altering the total production cross sections. 

More precisely, we will assume a KK mass scale $M_{\rm KK}=1.95$ TeV which leads to the 
following  masses for the various heavy neutral and charged resonances (in the mass basis)
\beq
M_{A_{KK}}=1.95~{\rm  TeV}, \  
M_{Z_{KK}}=M_{W_{KK}}=2~{\rm TeV}, \  
M_{Z'}=M_{W'}=1.96~{\rm  TeV}, \ 
\eeq
Assuming the equality of the two SU(2) gauge couplings $g_{L}=g_R$, we obtain the gauge boson
couplings to the SM heavy $(t,b)$ and light $(q)$ quarks by adopting the following $Q(c_{q_{L/R}})$ charges which comply with the data on the quark forward-backward asymmetries 
\beq  
Q(c_{t_R})=4 , \   Q(c_{t_L})=1.5 , \ Q(c_{q_L})= 0.3  , \ Q(c_{q_R})= - 0.2
\eeq
The couplings of the heavy vector bosons to light gauge (and Higgs) bosons are induced by 
mixing and the (sines of the) mixing angles have been given in a simplified approach in 
Ref.~\cite{KK-VV} that we closely follow and to which we refer for all details and notation. Numerically, they are are chosen to be   
\beq
s_1 =0.5 \, , \  s_{1c}=0.6 \, , \ s'=0.57 \, , \  
s_{0L}= - s_{0R}= - s_{01X}= s_{01}= 10^{-2}
\eeq
Using the numerical values quoted above, one can derive the total decay widths of the 5 resonances:   
\beq 
\Gamma_{A_{KK}}=68~{\rm  GeV}, \ 
\Gamma_{Z_{KK}}=98~{\rm  GeV}, \  
\Gamma_{Z'}= 95~{\rm  GeV}, \    
\Gamma_{W_{KK}}=68~{\rm  GeV}, \ 
\Gamma_{W'}= 210~{\rm  GeV}
\eeq
Note that while for the neutral states we assume that only decays into SM (heavy) fermions 
and SM vector and/or Higgs boson are kinematically accessible, in the case of the charged states, the additional channels $W_{KK}, W' \to tb'$ have been included.

\subsection*{3. The diboson excess}  \label{se:results}

We come now to the discussion of diboson production at the LHC and our tentative 
interpretation of some of the excesses observed by the ATLAS and CMS collaborations  
in our custodially protected warped extra--dimensional RS scenario.  

We have calculated the cross sections for the production processes $q\bar q 
\to W^+ W^-$,  $q\bar q' \to W^\pm Z$ and $q\bar q' \to W^\pm H$, including the $s$--channel
vector boson exchanges and the $t$--channel quark exchanges in the two first modes (in the third one, the Higgs couplings to light quarks are negligible). In the $WW$ case, the exchanged gauge bosons $V$  are the photon and $Z$ boson, their first KK excitations $A_{KK}, Z_{KK}$ and the $Z'$ boson; in the $WZ$ and $WH$ cases, the exchanged states are the $W$ and the heavier $W'$ and $W_{KK}$ resonances. The rates, where one should take into account the full interference, depend on the $V$ couplings to the initial $q \bar q$ pair and to the $W/Z$ or $H$ bosons as well as on the $V$ total widths. All these ingredients have been given in the previous section. Although the couplings of the heavy resonances to $W,Z,H$ states are induced by mixing and should in principle be small, the cross sections for longitudinal final states  will grow with powers of $M_{KK}^2/M_{W,Z,H}^2$ and, thus, will compensate for this suppression. In turn,  the $q\bar q \to ZZ$  process can be mediated only by $t$ and $u$--channel quark exchange as there is no coupling of the photon or $Z$ boson to $ZZ$ pairs and, according to the Landau--Yang theorem, heavy spin--one neutral vector bosons such as $Z'$ and $Z_{KK}$ cannot decay into two light ones.

\begin{figure}[!h]
\vspace{-.9cm}
\begin{picture}(400,200)
\hspace*{-12mm}
 \put(20,0){\includegraphics[width=6.3cm]{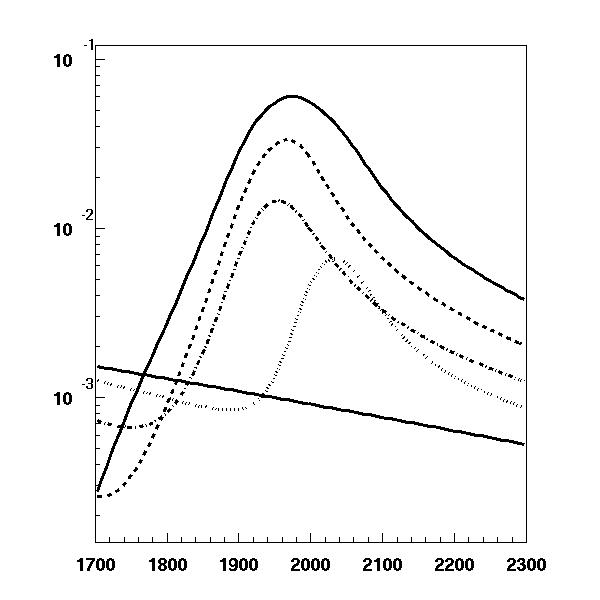}}\hspace*{3mm}
\put(170,0){\includegraphics[width=6.3cm]{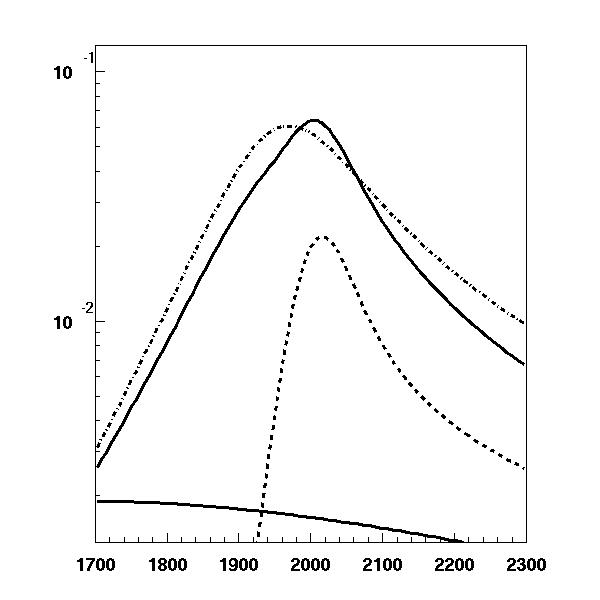}}\hspace*{3mm}
\put(320,0){\includegraphics[width=6.6cm]{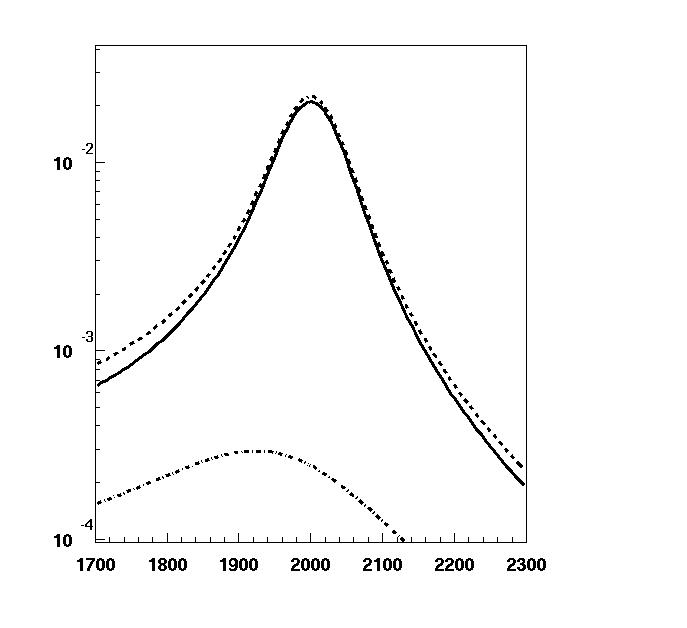}}
\put(25,+172){\color{blue} $\mathbf{ d\sigma/d M_{VV}~[fb/GeV]}$}
\put(175,-10){\color{blue}\bf diboson invariant mass [GeV]}
\put(126,145){\color{red}\bf WW}
\put(290,145){\color{red}\bf WZ}
\put(425,145){\color{red}\bf  WH}
\put(90,40){\color{blue}   SM+anom}
\put(256,28){\color{blue} SM+anom}
\put(39,145){\color{red} \bf tot}
\put(39,135){\color{red} $\mathbf{A_{KK}}$}
\put(39,125){\color{red} $\mathbf{Z'}$}
\put(39,115){\color{red} $\mathbf{Z_{KK}}$}
\put(197,145){\color{red} \bf tot}
\put(197,135){\color{red} $\mathbf{W'}$}
\put(197,125){\color{red} $\mathbf{W_{\hspace*{-1mm}KK}}$}
\put(357,145){\color{red} \bf tot}
\put(357,132){\color{red} $\mathbf{W_{\hspace*{-1mm}KK}}$}
\put(357,118){\color{red} $\mathbf{10W'}$}
\end{picture}
\vspace{3mm}
\caption{\small The differential cross sections (in fb/GeV) for the three processes $q\bar q \to W^+W^-$ (left), $q\bar q' \to W^\pm Z$ (center) and $q\bar q' \to W^\pm H$ (right) at the LHC with $\sqrt s=8$ TeV as functions of the diboson invariant masses (in GeV). The individual and total contributions of the various heavy resonances (ordered according 
to their importance) with masses close to 2 TeV are shown, together with the SM contributions including the ``anomalous" effects in the RS scenario.}\label{fig:VVxs}
\vspace*{-.1cm}
\end{figure}
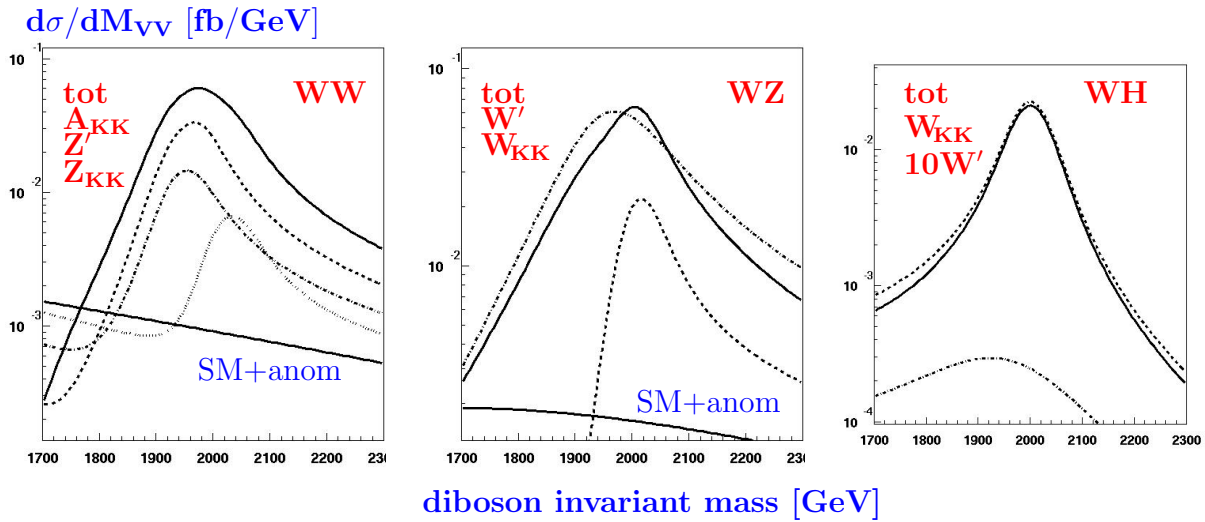

Figure~\ref{fig:VVxs} displays the differential cross sections at the LHC with $\sqrt s=8$ TeV as functions of the diboson invariant masses in the three processes, using MSTW parton distributions \cite{MSTW}. As can be seen, the small continuum contributions, which include the SM channels but with possibly significant new contributions at high  masses\footnote{These additional contributions result from the anomalous couplings among the SM gauge bosons induced by the RS scenario, which can be large as they are enhanced by powers of $M_{KK}^2/M_W^2$ \cite{first}.}, fall with the invariant mass of the diboson systems. However, there are significant excesses at a mass around 2 TeV which corresponds to the KK mass scale.  

The previous example shows that excesses in diboson final states due to resonances can be easily generated in the warped extra--dimensional scenario considered here. While the 
resonance mass needs to be fixed, the correct magnitude of the signal can be adjusted by simply tuning the various couplings of the KK states to SM particles. 
More specifically and channel by channel, the ATLAS and CMS data \cite{ATLAS-excess,CMS-excess} can be interpreted as follows. 

{\bf The WW mode:} The signal is obtained by considering the process $q\bar q \to V \to W^+ W^-$ with $V= A_{KK}, Z_{KK}$ and $Z'$ bosons.  Using the input KK masses and couplings given  previously, the ATLAS data with a $2.6\sigma$ excess in the $WW$ final state at an invariant mass of $\approx 2$ TeV can be reproduced. It turns out that the first KK excitation of the photon $A_{KK}$ is the main contributor to this particular final state. In addition, these three contributing resonances have moderate widths, below 100 GeV i.e. less that 5\% of the mass, one needs to take into account the detector mass resolution which is assumed here to be 4\%. As the three resonances have very close masses, with differences smaller than about 50 GeV and hence  the total decay widths, they are indistinguishable.   

Besides the SM contribution, there is also a pure QCD reducible background to the $pp\to WW \to $ jets topology: di-jet, $W/Z+$jet production etc... ATLAS has provided us with a formula that approximately describes this background from a parametrical adjustment of the data that have passed all selection and tagging requirements (it was found to be compatible both with simulated background events and several sidebands in the data).  The function includes all the relatively large uncertainties affecting them\footnote{This means that  the QCD background and the genuine $WW$ signal cannot be measured separately. While this has no impact for the observation of a resonance, it forbids measuring possible excesses in the $WW$ component due to possible anomalous couplings as also predicted by our scenario; see e.g. Ref.~\cite{first}.}. 

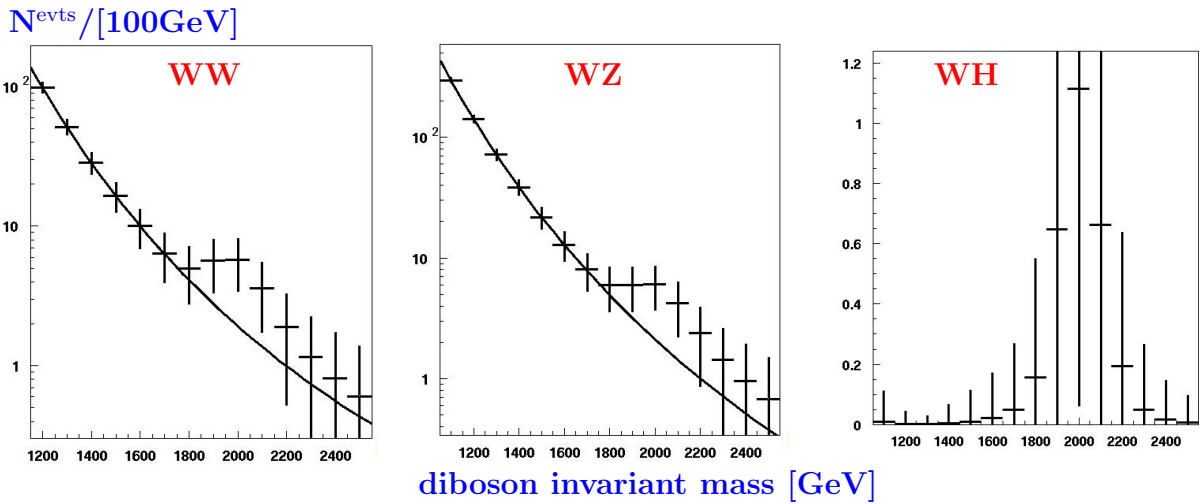
\begin{figure}[!ht]
\vspace{-1cm}
\begin{picture}(400,200)
\hspace*{-9mm}
 \put(20,0){\includegraphics[width=5.cm]{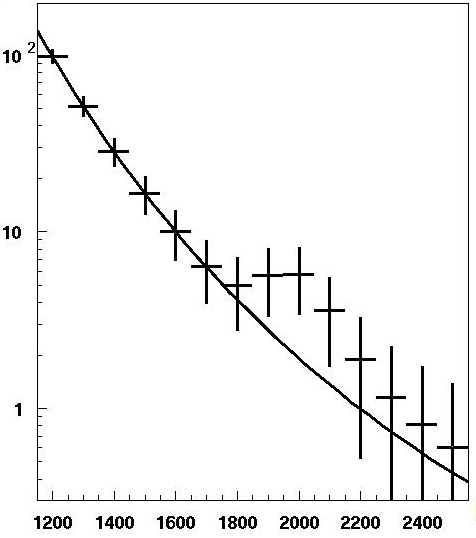}}\hspace*{2mm}
\put(170,0){\includegraphics[width=5.cm]{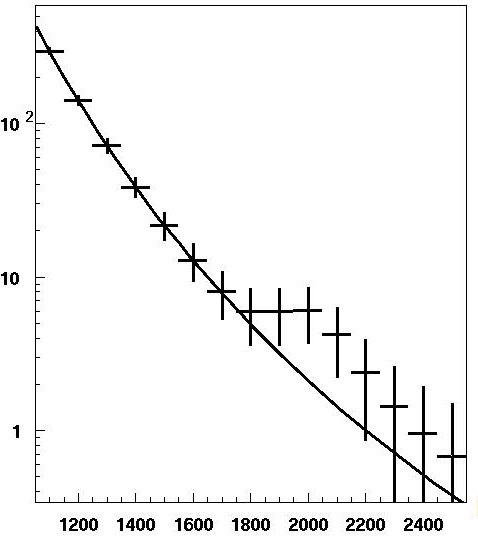}}\hspace*{-1mm}
\put(320,0){\includegraphics[width=6cm]{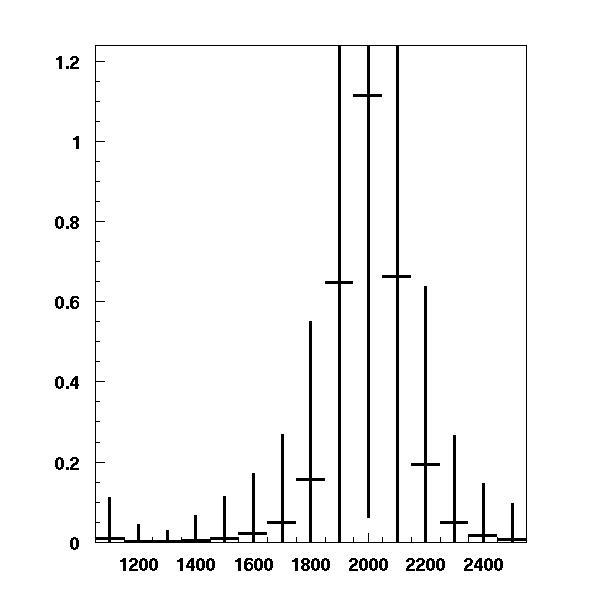}}
\put(20,+165){\color{blue}\bf  N$^{\rm evts}$/[100GeV]}
\put(175,-10){\color{blue}\bf  diboson invariant mass [GeV]}
\put(80,145){\color{red}\bf  WW}
\put(230,145){\color{red}\bf  WZ}
\put(370,145){\color{red}\bf  WH}
\end{picture}
\vspace{.1cm}
\caption{\small Expected mass distribution of dibosons at the LHC with $\sqrt s=8$ TeV and 20 fb$^{-1}$ data in the $WW$ (left), $WZ$ (center) and $WH$ (right) channels, assuming the efficiency and purity from the ATLAS (in $WW,WZ$) or the CMS (in $WH$) collaborations. Continuous lines are for the backgrounds and the bars are when adding the expected signals in our scenario.}\label{fig:VVbump}
\vspace{-.2cm}
\end{figure}

The left--hand side of Fig.~\ref{fig:VVbump} shows the expected mass distribution of the $pp \to W^+W^-$ process at the LHC at 8 TeV c.m. energy with 20 fb$^{-1}$ data,   assuming the efficiency and the purity given by ATLAS. The continuous line  corresponds to the predicted background and the simulated data, with their error bars, are obtained adding to this background the expected signal in our extra--dimensional scenario, with the numerical values of the parameters given before. As can be seen, the local $\approx 3\sigma$ excess observed by ATLAS at an invariant mass of $\approx 2$ TeV is reproduced within the statistical uncertainties. 

{\bf The WZ mode:} Once the various parameters of our scenario have been adjusted in order to fit the $WW$ data, the $WZ$ data and in particular the $\approx 3.4 \sigma$ ATLAS excess 
in this channel at a 2 TeV invariant mass can be straightforwardly explained in terms of 
$W'$ contributions to the process $q\bar q'\to WZ$ with the parameters given previously. The central plot of Fig.~\ref{fig:VVbump} shows the expected mass distribution of the $ZW$ final state at $\sqrt s=8$ TeV with 20 fb$^{-1}$ data, assuming the efficiency and purity given by the ATLAS collaboration. Again, the continuous line is for the QCD background while the simulated data are when the expected signal in our scenario is added on top of it. The excess is clearly visible. 

Compared to other approaches like in many GUT extensions of the SM for instance, our scenario has two interesting features. First, the $W_{KK}, W'$ and $Z_{KK},Z'$ states are predicted to have approximately the same mass, so we indeed have $M_{KK} \! \approx \! 2$ TeV for the four resonances. Second, the same mixing angles enter the couplings of the neutral and charged heavy states to $W/Z$ bosons and the rates for $q\bar q' \to WZ$ are fixed once the parameters entering $q\bar q \to WW$ are chosen.  Remarkably, they both turn out to be in agreement with the ATLAS data. The only freedom is, as already noticed in Ref.~\cite{KK-VV}, that additional decay modes into heavy quarks such as $W', W_{KK} \to t'b$ which affect the  resonance total widths, are possible. Since these new contributions cannot be predicted accurately, one can leave the total widths free to adjust the data more precisely. 

{\bf The ZZ mode:} In this case,  ATLAS observes an excess corresponding to a $2.9\sigma$ standard deviation. This excess is very difficult to explain in our context since, as
mentioned previously, heavy spin--1 neutral gauge bosons such as $Z_{KK}, Z'$, cannot decay into two almost massless neutral ones. One should thus assume that either the heavy resonance is the spin--two KK excitation of the graviton $G_{KK}$ with a mass that is close to $M_{KK} \approx 2$ TeV (although in the simplest scenarios the mass of $G_{KK}$ should be higher than this value). Another explanation  would be that the uncertainty in the measurement of the dijet mass could make one of the decaying $Z$ bosons resemble a $W$ or a $H$ boson, allowing the possibility to attribute the excess in our RS context to 
$WW,WZ$ or $WH$ production.

In fact, while ATLAS provides a good separation between the dibosons and the QCD background, there is a large overlap between $W$ and $Z$ selections and, hence, the existence of the three separate $WW, WZ$ and $ZZ$ channels is not certain, preventing a full comparison with our prediction. The only  meaningful attitude would be to sum the excesses in the three
different diboson modes. In doing so,  our scenario with the  parameters chosen above ideally predicts in the three 100 GeV most exciting mass bins around 2 TeV, a total of 9 signal events above the 9 background events which makes a total of 18 events. 

{\bf The WH mode:} The channel $pp \to WH$ with the subsequent decays $W \to \ell \nu$ and $H\to b\bar b$ has been searched for by CMS \cite{CMS-WH} and a $\approx 2.5$ standard deviation was found at an invariant mass of 2 TeV at which the SM background is negligible. In our scenario, the excess originates from $pp\to W_{KK} \to WH$ production. The right--hand side of Fig.~\ref{fig:VVbump} shows the expected signal in our scenario at $\sqrt s=8$ TeV with 20 fb$^{-1}$ data assuming the efficiency and the purity given by CMS. Note that the process $pp\to Z_{KK}, Z' \to ZH$ should be also observed at some stage but as the 
neutral cross section is smaller than that for the charge one and the leptonic $Z \to \ell^+ \ell^-$ branching rate is small, this neutral current process cannot be observed with the data collected at the previous LHC run.

We close this section by making two remarks. A first one is that $WW, WZ$ and $WH$ final states should  also be observed in  the semi-leptonic modes with similar sensitivities;  nevertheless, ATLAS observes no such signal, while CMS observes a $\approx 2.5\sigma$ effect in $WH$. A second remark is that if the $ZZ$ signal is due to a $\approx 2$ TeV mass KK graviton, the production would be initiated by gluon--gluon fusion and the significance of the signal could therefore increase when moving from 8 to 13 TeV energy. In fact, this would be a way to understand the origin of the signal without waiting for the observation of the much cleaner  leptonic final state which will require significantly more integrated luminosity.

\subsection*{4. Discussion and future prospects}

Let us now discuss the impact of our choice of parameters in the warped extra-dimension scenario that we consider, and in particular a KK mass scale  $M_{\rm KK} \simeq 2$ TeV, on the rates of the 125 GeV Higgs boson as measured at the LHC \cite{HiggsCombo}. The tree--level Higgs couplings to fermions and gauge bosons will be first affected by mixings between the SM fields and their KK excitations and the modification of the Higgs vacuum expectation value and, second, the loop-induced Higgs vertices will receive further contributions from exchanges of the KK towers of bosonic and fermionic modes as well as the 
custodians~\cite{KK-loop}. 

Within our RS framework, these effects can be parameterized in terms of two parameters besides $M_{\rm KK}$ \cite{LastNeubert}: the  size $kR_c$ of the extra dimension and the maximum absolute value $y_\star$ of the dimensionless complex Yukawa coupling. For fermion representations promoted to ${\rm SU(2)_L \times SU(2)_R \times U(1)_X}$ multiplets with equal ${\rm SU(2)_R}$ and ${\rm SU(2)_L}$ gauge coupling constants, the predictions for the Higgs production and decay rates were calculated in Ref.~\cite{LastNeubert}. It was shown that for reasonable $kR_c$ and $y_\star$ values and in two different scenarios, one of a Higgs field localized towards the TeV--brane but with a narrow width profile and another of a purely brane--localized Higgs field, one needs $M_{\rm KK}$ values beyond a few TeV and in any case, $M_{\rm KK} \gsim 2.5$ TeV, in order to cope with the Higgs data.  



A detailed analysis of the Higgs production and decay rates in the RS scenario considered here is beyond the scope of this note. Nevertheless, we believe that there is a way to cope with the LHC data on the Higgs signal strengths $\mu_{XX} = \sigma(pp  \to H \to XX)/\sigma(pp \rightarrow H \to XX)|_{\rm SM}$  for the dominant detection channels,  $H \to \gamma \gamma, ZZ^* \to 4\ell,  WW^* \to 2\ell2 \nu, b \bar b$ and $ \tau^+ \tau^-$. The reasons behind this optimism are the following.  

First in their combined analyses of the Higgs signals, the  ATLAS and CMS collaborations assumed that all uncertainties can be treated as Gaussian which is not entirely correct as the theoretical uncertainties, which are at the level of 10--15\% and  have been underestimated by the experiments, should be treated as a bias; see Ref.~\cite{DTHiggs} for detailed discussions. The total uncertainties on the signal strengths are thus larger and, at the 2$\sigma$ level, one could still allow for a deviation of order of 50\% that a mass scale of $M_{\rm KK} \simeq 2$ TeV can generate on the most precisely measured  $\mu_{WW}, 
\mu_{ZZ}$ and $\mu_{\gamma\gamma}$ signal strengths.  

Second, one could include the effects of the new quarks $t'$ and $b'$ that we do not completely specify here as we are mainly focusing on the bosonic sector (the effects of the fermionic KK excitations have been included in the analysis of Ref.~\cite{LastNeubert} in 
an approximate way but not the ones of the ``custodians"). They could generate global modifications to the loop induced processes, such as the dominant $gg\to H$ production mechanism and the precisely measured $H\to \gamma\gamma$ decay mode. These 
new partners will also alter the tree--level $Ht\bar t$ (and $H b \bar b$) couplings through fermion mixing. The combined effects could even lead to  $\mu_{XX}$ values that are more compatible with data\footnote{An example of such a situation has been recently given in Ref.~\cite{ttH-VLQ} in the context of vector--like top and bottom quark partners which contribute to the $Hgg$ and $H\gamma\gamma$ loop--induced vertices to make the signal strengths in the main channels closer to their experimental values. In addition, these new fermions could explain the observed excess in the $pp\to t\bar t H$ associated production channel \cite{HiggsCombo}.}.  

Finally, a third point is that these indirect constraints are also subject to uncertainties   from  higher dimensional non-renormalisable operators originating from the ultra-violet completion of the model. The latter can potentially lead to large (and hopefully compensating) effects; see for instance, the recent analysis of Ref.~\cite{higher-dimension} in a similar context.     

In fact, a similar problem occurs when addressing the indirect constraints from electroweak precision data. While the direct radiative corrections to the $Z$ partial decay width into  bottom quarks are taken care of by our choice of parameters that fit the asymmetries $A_{FB}^{t,b}$, there are too large contributions in the so--called oblique corrections that affect the $W$ boson mass and the effective mixing angle $\sin^2\theta_W$ at high orders. Indeed, even under the hypothesis of a custodial symmetry ${\rm SU(2)_L \times SU(2)_R \times U(1)_X}$ gauged in the bulk that should in principle allow for some protection, analyses of oblique corrections lead to a lower bound of a few TeV on the mass $M_{\rm KK}$ in the simplest realisations \cite{RSWidth}.

In conclusion, we will consider the KK resonance mass scale $M_{\rm KK} \simeq 2$~TeV to be viable despite the  potentially problematic corrections to the Higgs signal rates and the  electroweak precision data as they can be alleviated by introducing additional degrees of freedom or new contributions. In some sense, we adopt the spirit of a bottom-up approach and consider that the direct signal of new physics, like the production of new gauge bosons, should probably be taken more seriously than the indirect constraints from virtual heavy particle exchanges. 

Another important remark that should be made is that small excesses, of about two standard deviations or less,  have also been observed by the experimental collaborations in the two production processes $pp\to t\bar t H$ \cite{HiggsCombo} and $pp\to t\bar t W$ \cite{ttW} and it is tempting to interpret them in our warped extra--dimensional scenario. The simplest interpretation would a KK gluon with a  mass of $\approx 2$ TeV that is produced with a significant cross section and which decays into the heavy top and bottom quark partners that are also predicted in the scenario. The topologies $pp\to g_{KK} \to
t't'$ and $t't$, with a significant $t' \to tH$ decay branching ratio, would contribute to the $t\bar t H$ final state. The new $g_{KK} \to b'b'$ production mode,  with the $b' \to tW$ decay, would lead to a $4W$ topology that would give like--sign leptons, missing energy and jet activity in the final state and hence, explain the $ttW$ excess. We have checked that, indeed, a KK gluon with $M_{KK} \approx 2$ TeV decaying into heavy quarks with masses above $m_{Q'} \gsim 0.9$ TeV \cite{PDG} can produce the observed excesses without being in conflict with other experimental  data. For instance, a 2 TeV KK gluon would mainly decay into $t\bar t$ pairs and would be in principle excluded by searches of di-top resonances \cite{PDG}. However the present LHC limit $M_{KK} \gsim 2$ TeV hold only if the resonance is narrow, 
$\Gamma/m \lsim 10\%$, which is not the case of our $g_{KK}$  whose total width is much 
larger, especially if the extra channels $g_{KK}\to t't', tt', b'b', bb'$ are open. 

Before summarizing, let us also briefly discuss the implication of upgrading the LHC c.m. energy from 8 TeV to 13 TeV and the prospects for strengthening the diboson signal with the  expected $\approx  4$ fb$^{-1}$ data that has been collected this year. For a 2 TeV resonance, there is an increase of a factor of $\approx 6$ in $q\bar q$ luminosity when moving from $\sqrt s\!= \!8$ TeV to $\sqrt s\!= \! 13$ TeV \cite{Luminosity} and, therefore, with the 4 fb$^{-1}$ data sample collected so far, the number of signal events should be of the same order of magnitude as the present one. 

For the QCD background, a precise statement cannot be made at this stage as, for instance, the ATLAS collaboration uses the data itself to normalize it. A precise calculation of the major backgrounds is beyond our scope here, but approximately one expects the following pattern. If the main source of background is originating, as it seems to be the case, from $qq , q\bar q$ and $qg$ initiated processes,  one should have the same significance for the signal as in the $\sqrt s \! =\! 7\! +\! 8$ TeV run  with the 4 fb$^{-1}$ data sample. If the main source  is instead due $gg$ fusion,  for which the parton luminosity increases faster with a factor of $\approx 18$ from $\sqrt s \! = \! 8$ to $\sqrt s\! =\! 13$ TeV for a 2 TeV invariant mass \cite{Luminosity}, the significance of the diboson signal would be significantly reduced (probably below $\approx 1.5\sigma$ for the 4 fb$^{-1}$ data).  We are thus anxiously waiting for the analyses of these excesses at the current LHC run at $\sqrt s=13$ TeV that should be released soon. 

\subsection*{5. Conclusions}

We have considered the diboson excesses observed by the LHC experiments in both the $WW,WZ$ and $WH$ production channels and interpreted them in terms of the production  of heavy spin--one resonances:  the Kaluza--Klein excitations of the electroweak gauge bosons in the context of a custodially protected model of warped extra space--time dimensions. We have focused our attention on scenarios that also address two anomalies in the heavy quark sector of the SM: the bottom and (to a lesser extent) top quark forward--backward asymmetries as measured respectively at LEP and at the Tevatron. 

We have indeed found a set of parameters of the model that nicely fits the ATLAS and CMS diboson data, except for the excess in the $ZZ$ channel that is very difficult to explain unless one assumes a comparable mass for KK gravitons and gauge bosons or a miss-measurement of the dijet invariant mass which would make that one of the $Z$ boson is actually either a $W$ or a Higgs boson. The price to pay for this scenario with a resonance mass scale $M_{KK}=2$ TeV  is some tension with the LHC Higgs and electroweak precision data, but which can be alleviated by allowing for additional contributions from other sectors of the model such as e.g. heavy top and bottom quark
partners. 
 
While the approximately 4 fb$^{-1}$ data collected this year at $\sqrt s\! =\! 13$ TeV should shed more light into this diboson excess, a firm conclusion can only be reached next year
when the collected data set will exceed the 20 fb$^{-1}$ level. In the very exciting eventuality that this phenomenon will persist with this larger data set, it would be necessary to refine the ``preliminary" analysis presented in this note. In particular, we would need to make a more adequate choice of the parameters of this warped extra--dimensional scenario in order to fit more accurately the experimental data and to evade in a more serious way the constraints from the Higgs measurements at the LHC and the precision electroweak data. All these issues will be postponed to  a forthcoming publication. \smallskip 

\noindent {\bf Acknowledgements:} Discussions with and suggestions from Kaustubh Agashe are greatfully acknowledged. A.D. would like to thank the CERN Theory Unit for hospitality. The work is supported by the ERC advanced grant {Higgs@LHC}. G.M. is supported by the Institut Universitaire de France  and the European Union FP7 ITN ``Invisibles".

\end{document}